\documentclass[12pt]{iopart}
\usepackage{cite}
\usepackage{graphicx}
 \usepackage{amssymb}
 \usepackage{subfigure}
 \usepackage{textcomp}
 \usepackage{color}
 \usepackage{amsfonts}
 \usepackage{bbold}
 \usepackage{dsfont}
 \usepackage{epsfig}
\usepackage{dcolumn}
\DeclareGraphicsExtensions{.eps}
\graphicspath{{figures/}}
\usepackage{subfigure}
\usepackage{epstopdf}

\begin{document}

\title{Spin transport in higher $n$-acene molecules}

\author{R.~Pilevarshahri$^1$, I.~Rungger$^2$, T.~Archer$^2$ , S.~Sanvito$^2$ and N.~Shahtahmasebi$^1$}

\address{1)Physics Dep., Faculty of Sciences, Ferdowsi University of Mashhad, Iran ~2)School of Physics and CRANN, Trinity College, Dublin 2, Ireland}
\eads{\mailto{pilevarr@tcd.ie}, \mailto{runggeri@tcd.ie}}
\date{\today} 
\begin{abstract}
We investigate the spin transport properties of molecules belonging to the acenes series by using density functional theory 
combined with the non-equilibrium Green's function approach to electronic transport. While short acenes are found 
to be non-magnetic, molecules comprising more than nine acene rings have a spin-polarized ground state. In their gas phase
these have a singlet total spin configuration, since the two unpaired electrons occupying the doubly degenerate
highest molecular orbital are antiferromagnetically coupled to each other. Such an orbital degeneracy is however lifted 
once the molecule is attached asymmetrically to Au electrodes via thiol linkers, leading to a fractional magnetic moment. 
In this situation the system Au/$n$-acene/Au can act as an efficient spin-filter with interesting applications in the emerging 
field of organic spintronics.

\end{abstract}

 \maketitle
 \section{Introduction}
\begin{figure}[ht]
          \begin{centering}
         \includegraphics[width=0.2\linewidth]{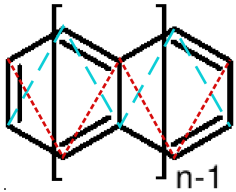}
        \caption{\label{acene}\small{(Colour on line) Schematic illustration of the $n$-Acene series. The dashed (light blue) and
        dotted (red) line denote the symmetry of the two degenerate HOMOs.}}
\end{centering}
\end{figure}

%
Molecules of the $n$-Acenes family, C$_{4n+2}$H$_{2n+4}$, are an important class of organic compounds formed from polycyclic aromatic hydrocarbons 
consisting of linearly fused benzene rings. These molecules, schematically shown in figure~\ref{acene}, can be also viewed as small hydrogen 
terminated graphene nano-ribbons. The interest in their electronic properties has grown enormously in the past few years~\cite{Chun,Kaur,Mondal,Payne,Antony,Tonshoff,Bendikov,Bendikov-c,Santos,Houk,Hachmann,jiang,Hajgato,Zade,clar,Bettinger,Jurchesco,Qu,Angliker,Nijegorodov,Voz,Yanagisawa,Amrani,Diao}. 
The smallest acenes, benzene ($n=1$) and naphthalene ($n=2$), are well known aromatic compounds, while the intermediate ones, 
tetracene ($n=4$) and pentacene ($n=5$), are semiconductor materials in their solid phase. These are commonly used in organic field 
effect transistors~\cite{Antony,Park}, organic light emitting diodes~\cite{Jang,Wolak} and organic photovoltaic cells~\cite{Rand}, because of 
their high charge carrier mobility. For pentacene single crystals a value of 35~$\mathrm{cm}^{2}\mathrm{V}^{-1}\mathrm{s}^{-1}$ was
reported~\cite{Jurchesco}, which is among the highest mobility of all organic semiconductors.  As we move to still larger acenes, which we 
will refer here to as higher acenes, such as heptacene ($n=7$), octacene ($n=8$), nonacene ($n=9$) and decacene ($n=10$), the structural 
stability is reduced. Those molecules are in fact quite reactive and the experimental literature become less rich.

Theoretical studies predict that the higher acenes are good candidates for realizing $p$-type magnetism, i.e. magnetism without 
ions presenting partially filled $d$ or $f$ shells~\cite{Bendikov,Bendikov-c,jiang,Houk,Hachmann,Santos}. 
If the larger magnetic acenes can be stabilized, their magnetism coupled with the long spin-diffusion length will make this material 
class extremely appealing for the next generation of organic spintronics devices \cite{Stefano,Stefano2,dediu1,WW1,Greg}.

The smaller acenes have a non-magnetic (diamagnetic) ground state. For the higher acenes calculations predict the edges 
of the molecule to be spin polarized~\cite{Bendikov,Bendikov-c,Santos,jiang,Hachmann,Qu}. In this case there are two degenerate 
highest occupied molecular orbitals (HOMOs), each of them populated by a single electron. One of the HOMOs is localized mainly 
on the hydrogenated carbon atoms at one edge of the molecule, while the other HOMO is localized on the opposite edge. These 
two degenerate molecular orbitals extend in a zigzag-like symmetry along the length of the molecule as indicated by the dashed 
lines in Fig~\ref{acene}. These states are similar in nature to the edge states found in graphene nano-ribbons~\cite{graphenepaper}. 
The lowest unoccupied molecular orbitals (LUMOs) are the corresponding empty spin-split orbitals, so that the HOMO-LUMO gap 
is an exchange split gap. Recent calculations predicted a spin singlet ground state for the higher acenes \cite{Bendikov,Bendikov-c,jiang, Hachmann,Qu} 
as a result of the antiferromagnetic coupling between the two orbitally degenerate HOMOs. This contrasts earlier predictions of 
a ferromagnetic (triplet) ground state \cite{Angliker,Nijegorodov,Houk}. 

Since small acenes are diamagnetic and the higher ones are predicted magnetic, there is a critical length for the onset of
magnetism. This is generally accepted to occur for $n$ between 6 and 8, but there is no consensus on the precise value. 
Houk {\textit{et al.}}~\cite{Houk} used the B3LYP functional and found that for more than 8 acene rings a triplet ground state 
is lowest in energy. Bendikov {\textit{et al.}}~\cite{Bendikov,Bendikov-c} performed calculations at the UB3LYP level and 
observed a magnetic ground state already for 6 acene rings. Finally Jiang {\textit{et al.}}~\cite{jiang} found the transition to a magnetic 
ground state for more than 7 benzene rings.  

From the experimental point of view, a significant amount of work has been devoted to the study of electron transport 
through pentacene~\cite{Voz,Yanagisawa,Amrani,Diao}. The same material has also been used as a medium for spin-polarized 
transport in composite junctions \cite{Shimada} with some evidence for spin-injection \cite{Ikegami}. Pentace itself however is
not magnetic. For this reason significant new effort is being made to synthesize the higher acenes~\cite{Chun,Kaur,Antony,Payne,Tonshoff, Mondal,Bettinger,Zade}. 
After much skepticism about its stability~\cite{clar}, heptacene was finally synthesized in 2006, and subsequent analysis has shown 
that the ground state is non magnetic~\cite{Mondal,Bettinger}. More recently Tonshoff and Bettinger synthesized octacene and 
nonacene~\cite{Tonshoff}, and importantly there is spectroscopical evidence for anti-ferromagnetism in nonacene~\cite{Tonshoff}.

In this work we investigate both the ground state electronic structure and the electronic transport properties of the higher acenes. 
In particular we focus our attention on pentacene and decacene, in order to compare results for a non-magnetic molecule, such as 
pentacene, to those of a magnetic one, such as decacene.

\section{Methods} 

The electronic structure is calculated by using spin polarized density functional theory (DFT), as implemented in the SIESTA code~\cite{siesta}. 
The transport properties are then obtained with the non-equilibrium Green's function (NEGF) code SMEAGOL~\cite{smeagol1,smeagol2,senepaper}, which shares
the Kohn-Sham Hamiltonian with SIESTA. The LDA functional~\cite{LDA} is used throughout this work. A double-$\zeta$ (DZ) basis was used to 
describe the electronic structure of the valence electrons of the molecule, while we use an $s$ only single-$\zeta$ basis set for the gold electrodes. 
This has been extensively tested in the past~\cite{magmol,bdtcntwater} and offers a good compromise between accuracy and computational demands. 
A real space grid with an equivalent plane wave cutoff of 300 Ry is used to sample the electronic density. For the isolated molecule periodic images 
are separated by a minimum of 10~\AA, which is found to be sufficient to prevent interaction between the image molecules. For all the molecules 
structural relaxation is performed by standard conjugate gradient to a force tolerance of 0.01~eV/\AA. 

The transport junction is constructed by connecting the molecules to a Au fcc (111) surface via sulfur anchoring groups with a S-Au bond length 
of ${1.9}$~\AA \cite{cormacalkane}. We use a $k$-point mesh with a cutoff of 9~\AA~perpendicular to the transport direction. Due to the presence 
of many local minima, which are found for the different local magnetic alignments, it was necessary to initialize the calculations with several different 
spin configurations for both the electronic structure and the transport calculations in order to ascertain the global ground state.
\section{Results}

\subsection{Electronic structure of $n$-Acene molecules: gas phase}

\begin{figure}[ht]
  \centering
  \subfigure[\small{}]{\label{pentacene_up}
  \includegraphics[width=0.25\linewidth]{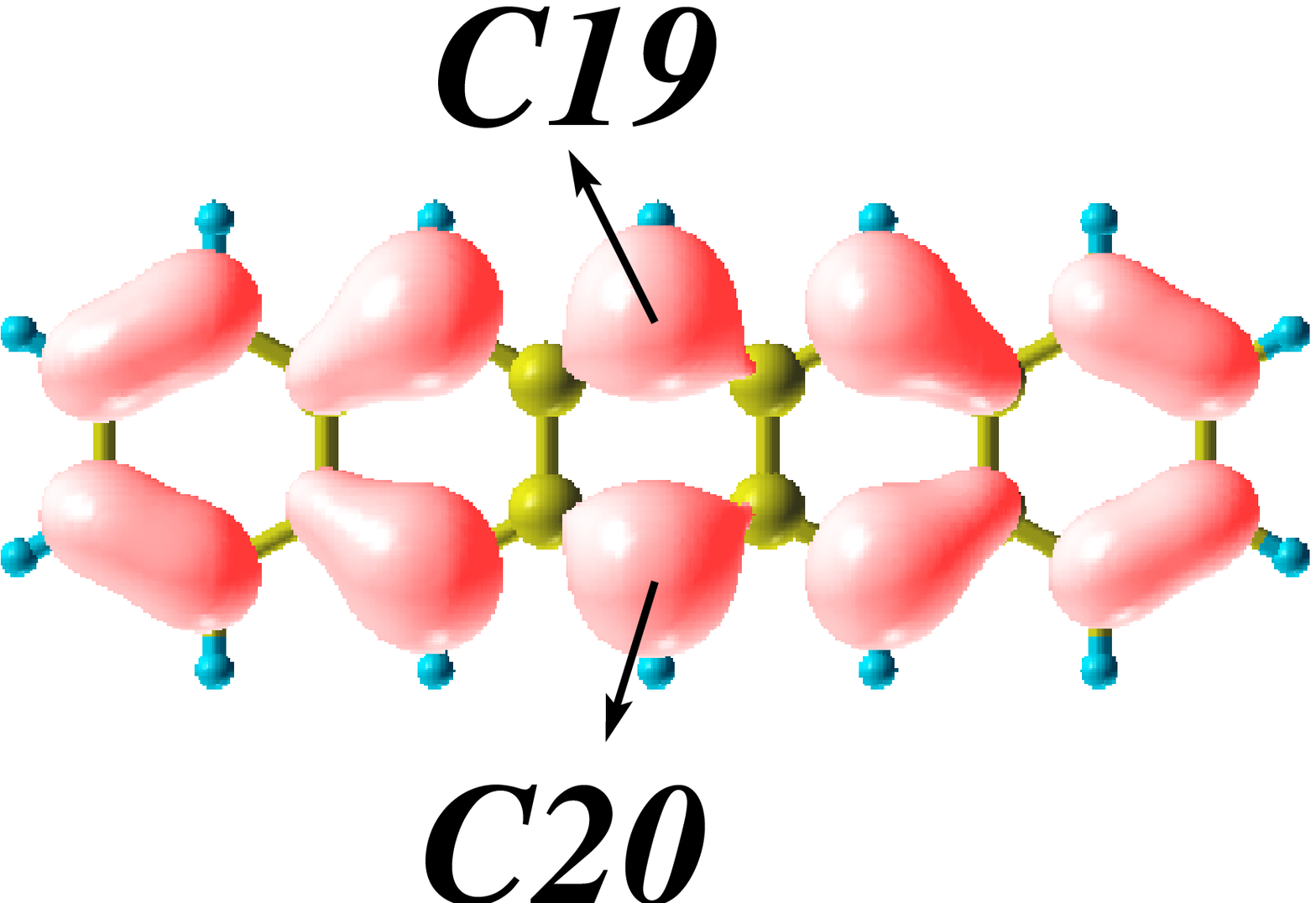}}
  \hspace{0cm}
  \subfigure[\small{}]{\label{decacene_up}
  \includegraphics[width=0.50\linewidth]{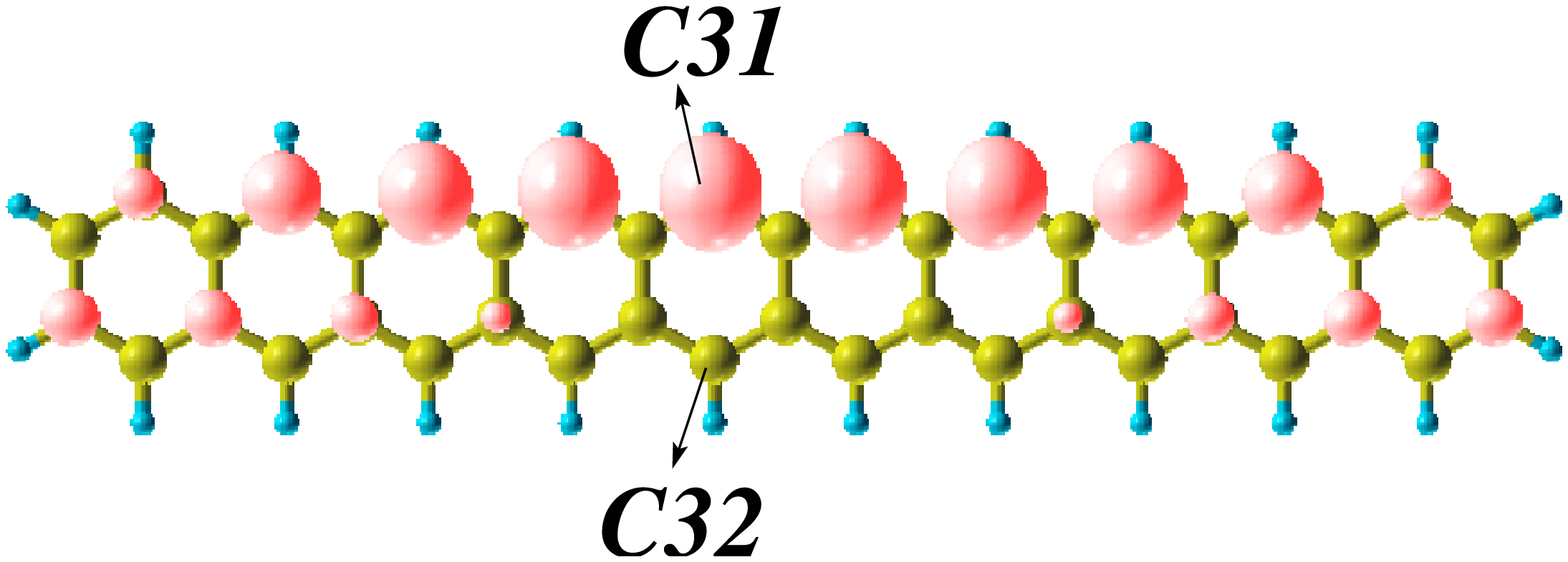}}
  \hspace{0cm}
  \subfigure[\small{}]{\label{pentacene_down}
  \includegraphics[width=0.25\linewidth]{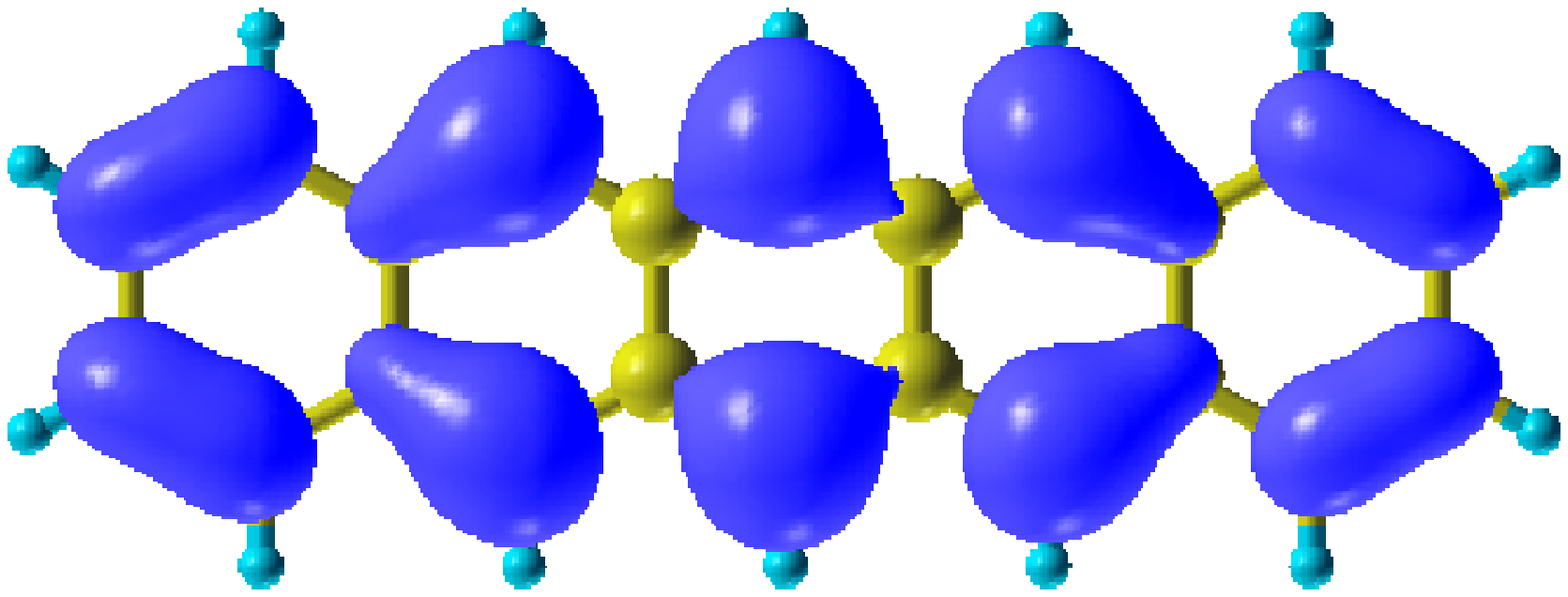}}
  \hspace{0cm}
  \subfigure[\small{}]{\label{decacene_down}
  \includegraphics[width=0.50\linewidth]{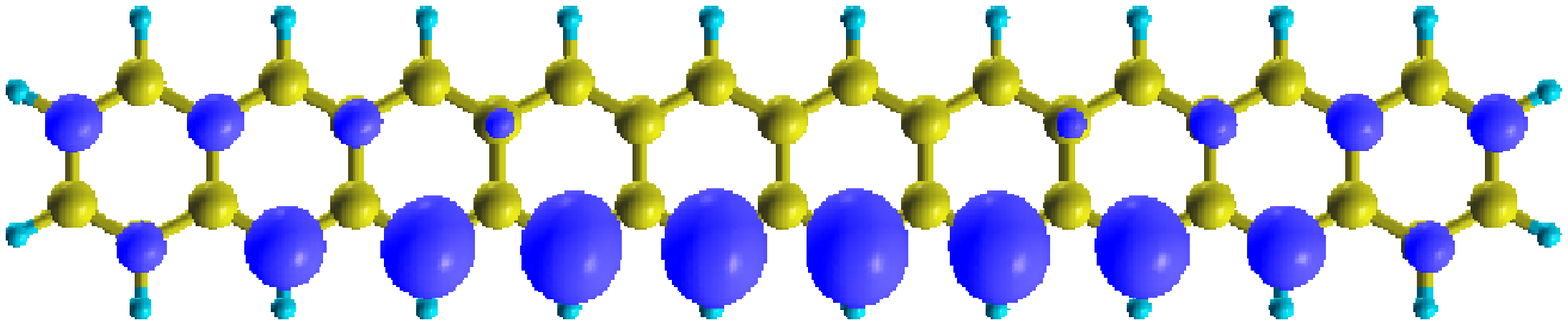}}
  \hspace{0cm}
  \caption{\small{(Colour on line) Spin-resolved electron density of the HOMO of both pentacene and decacene. The various panels are respectively for: 
  (a) pentacene majority spins, (b) decacene majority spins, (c) pentacene minority spins, (d) decacene minority spins. In the molecule structure the large
  light grey (yellow) spheres are for C and the small grey (light blue) ones are H.}} 
  \label{isolated}
\end{figure}

We calculate the electronic structure of the isolated $n$-acenes for the molecules of different lengths ranging from 5 acene rings ($n=5$, pentacene) 
up to 10 ($n=10$, decacene). A non magnetic ground state is found for $n\le8$, whereas nonacene and decacene present a spin polarized ground state. 
For these the atoms at the upper and lower edges of the molecules have an equal and opposite magnetic moment, corresponding to an anti-ferromagnetic 
local spin arrangement (see figure~\ref{isolated}). This is consistent with previous results on higher acenes~\cite{Bendikov,jiang}, and it is similar to the 
edge-state magnetism predicted for graphene nano-ribbons~\cite{graphenepaper}. In decacene the total energy for the anti-ferromagnetic phase is 
$\Delta E=40$~meV smaller than that of the ferromagnetic one and $\Delta E=110$~meV smaller than that of a non-spin polarized solution.

\begin{figure}[ht]
  \begin{centering}
  \includegraphics[width=0.7\linewidth,clip=true]{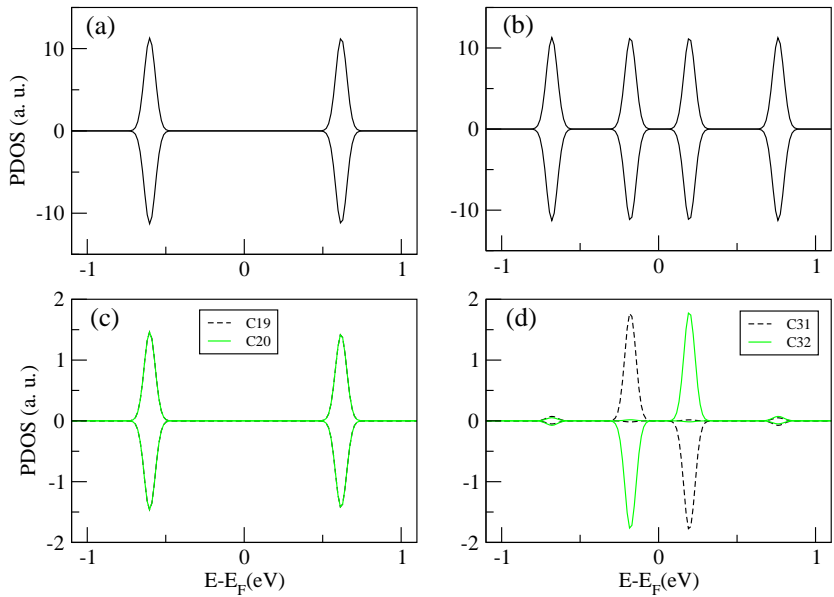}
  \caption{\label{PDOS_M_T}\small{Projected density of states (PDOS) over selected atoms: (a) all the C atoms of pentacene, (b) all the C
  atoms of decacene, (c) middle two C atoms of pentacene, (d) middle two C atoms of decacene. Positive PDOS are for the majority spins and
  the negative are for the minority.}}
  \end{centering}
\end{figure}

The total spin-resolved density of states (DOS) projected over all the carbon atoms is shown in figure~\ref{PDOS_M_T}(a) for pentacene and in 
figure \ref{PDOS_M_T}(b) for decacene. In both cases there is no net spin-polarization. However, while for pentacene all the molecular orbitals
are spin-degenerate, in decacene spin-symmetry is locally lifted. This can be appreciated by looking at the projected density of states (PDOS) on 
the two carbons atoms located in the middle of molecules at opposite edges [in the case of decacene these are the atoms labelled as C31 and 
C32 in figure~\ref{decacene_up}]. The PDOS are shown in figure~\ref{PDOS_M_T}(c) and ~\ref{PDOS_M_T}(d) for pentacene and decacene, respectively. 
It is then clear that for decacene the majority spins are located on the upper edge, while the minority ones on the lower edge. This means that,
although decacene does not provide a net spin polarization over the entire molecule, it is locally magnetic. It is then expected that a net magnetic 
moment can be created by breaking the molecule axial symmetry. In the following sections we will demonstrate that this can indeed be achieved 
when attaching the molecule asymmetrically to metallic electrodes.

\begin{figure}[ht]
  \begin{centering}
  \includegraphics[width=0.6\linewidth,clip=true]{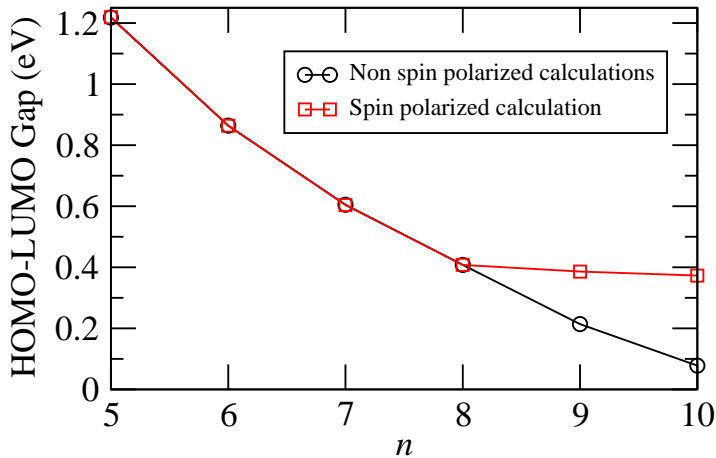}
  \caption{\label{gap}\small{HOMO-LUMO gap calculated from the Kohn-Sham eigenvalues for $n$-acenes as a function of $n$.}}
\end{centering}
\end{figure}

We then calculate the HOMO-LUMO gap for all the acenes ranging from pentacene to decacene (see figure~\ref{gap}). As the number of benzene rings 
increases, the HOMO-LUMO gap is reduced. Interestingly the dependence of the HOMO-LUMO gap on the acene length changes as $n>8$, i.e. as
the molecules develop a spin-polarized ground state. In fact for the magnetic molecule the gap is an exchange gap between the spin split molecular 
orbitals with the same orbital symmetry. Such an exchange gap is not strongly length dependent and it is expected to saturate to a constant value
characteristic of the infinite graphene ribbon. Finally we note that, if a non-magnetic solution is forced one will expect a gap closure for sufficiently
long molecules. 

It is important to remark at this point that the limitations of the LDA exchange correlation functional make our calculated HOMO-LUMO gap 
smaller than the experimental one. The experimental gap of pentacene in the gas phase is found to be 5.54~eV \cite{Dolomatov}, whereas 
our calculations report a value of only about 1.2~eV. Certainly one should not expect the Kohn-Sham HOMO-LUMO gap to reproduce the 
true excitation gap of the molecule. Still in the NEGF scheme the Kohn-Sham eigenvalues are effectively used as single particle levels
for the transport, and one should then ask whether the LDA description is sufficient \cite{Cormac}. Notably when a molecule is attached to 
a to metallic surface its ionization potential (IP) is usually reduced due to large correlation effects. In \cite{neatongw} the 
HOMO-LUMO gap of pentacene on graphite is calculated by the GW method to be about 2.88 eV. We therefore expect that our LDA 
HOMO-LUMO gap for pentacene on the surface is underestimated by about a factor 2 with respect to experiments. In fact, experimental 
data \cite{Koch} return a HOMO-LUMO gap of about 2.2~eV for pentacene thin films on Au. There are no experimental data on the decacene 
IP, however we expect a deviation from experiments similar to the one of pentacene. 

\subsection{Transport properties}

\begin{figure}[ht]
  \centering
  \subfigure[\small{}]{\label{meta}
  \includegraphics[width=0.55\linewidth,clip=true]{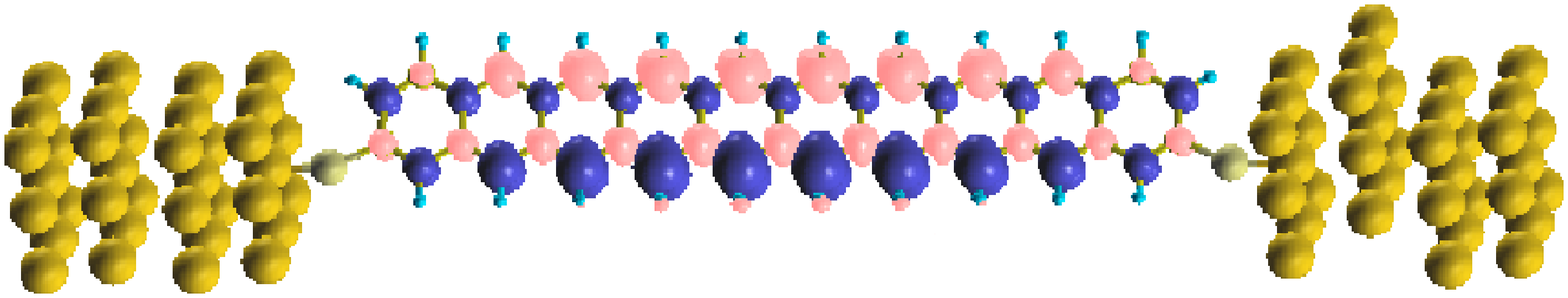}}
  \hspace{0cm}
  \subfigure[\small{}]{\label{para}
  \includegraphics[width=0.55\linewidth,clip=true]{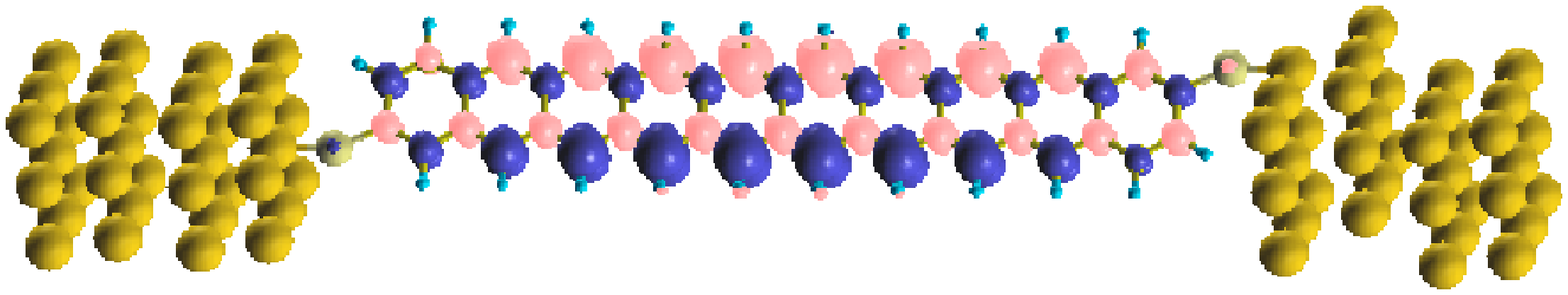}}
  \hspace{0cm}
  \subfigure[\small{}]{\label{3}
  \includegraphics[width=0.55\linewidth,clip=true]{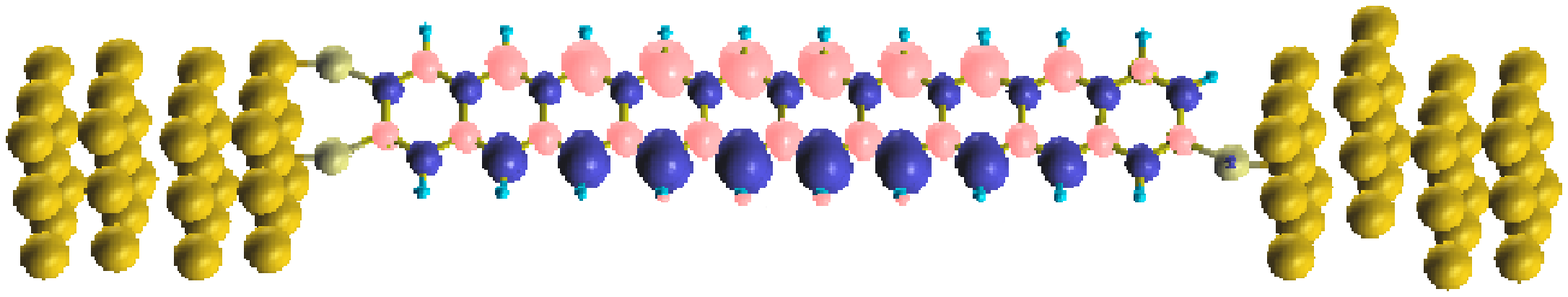}}
  \hspace{0cm}
  \subfigure[\small{}]{\label{4}
  \includegraphics[width=0.55\linewidth,clip=true]{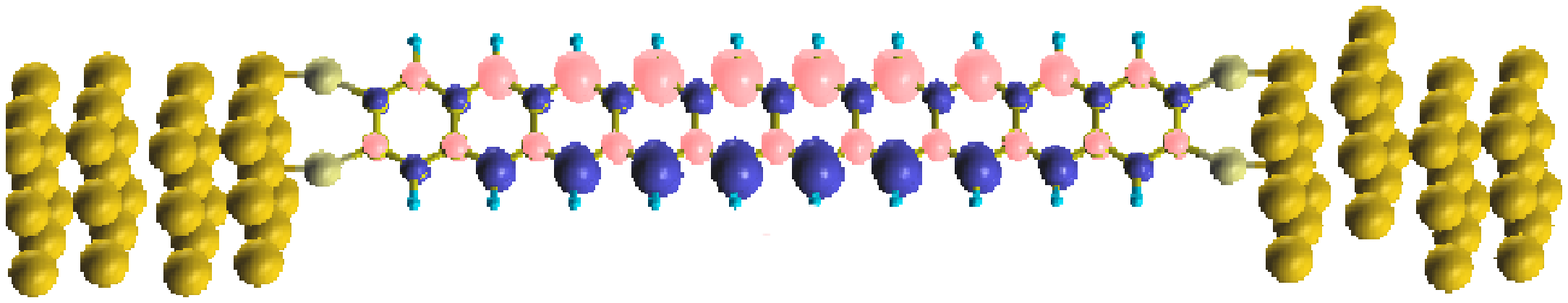}}
  \hspace{0cm}
\caption{\small{(Colour on line) Isosurfaces of spin density magnetization for the various anchoring geometry of decacene attached to Au via a 
thiol group: a) two links, meta; b) two links, para; c) three links d) four links. Positive isosurfaces are represented in light grey (pink) and negative
ones in dark grey (blue).}}
\label{homo_deca}
\end{figure}

We now investigate the electronic structure and the transport properties of the molecules attached to Au electrodes. We use thiol groups to bind 
the acenes to the Au surface and consider four different binding geometries. These are shown for decacene in~figure \ref{homo_deca}. 
Panel (a) shows the meta configuration, in which only the two ends of the lower edge are connected to Au. figure \ref{homo_deca}(b) 
shows the para configuration, where the lower edge on the left side and the upper edge on the right side are connected to the electrodes. 
figure \ref{homo_deca}(c) corresponds to the meta configuration with an additional S linker on the left side upper edge, and in 
figure \ref{homo_deca}(d) there are 4 S linkers. 

The ground state of the isolated thiolated molecules is anti-ferromagnetic for decacene, and non-magnetic for pentacene. This demonstrates
that the electronic structure of the isolated molecules is affected little by the attachment of the S-H end-groups. We typically run multiple 
simulations for each geometries by initializing the density matrix in different spin states. In the case of the complete junction (electrodes
plus molecule) we find always an identical final spin configuration regardless of the initialization, so that any other metastable solution
cannot be stabilized during the self-consistent cycle. In particular the calculations for pentacene all converge to a non-magnetic solution,
while those for decacene to a spin-polarized one. In this second case the coupling to the electrodes determines the final magnetic state. 

When the coupling is asymmetric [figures \ref{homo_deca}(a) and \ref{homo_deca}(c)] the system develops a significant fractional magnetic 
moment, while no moment is found for the symmetric geometries [figures \ref{homo_deca}(b) and \ref{homo_deca}(d)].
In general the coupling to the electrodes leads to level broadening and consequently to charge transfer from the molecule HOMO to the Au. 
Thus, whereas the majority HOMO is always fully filled, the minority one may have partial occupation and the total spin-polarization may be 
non-zero. The fact that different density matrix initializations result in the same solution indicates that, at variance with the isolated molecule 
which possess different local energy minima corresponding to different magnetic states, when the molecule is attached to Au there is only 
one robust global minimum. For the meta configuration the energy of this single global minimum is about $100$~meV smaller than that of the 
non-spin polarized solution.
figure \ref{homo_deca} shows the isosurfaces associated to the spin-density (the charge density difference between majority and minority spins) of decacene 
for the four geometries investigated. From the figure, one can notice the general antiferromagnetic coupling between the spins of the two
edge states, a characteristic feature of the molecule in its gas phase, which is largely preserved when the molecular junction is made. The coupling to
the electrodes in some cases breaks the axial symmetry of the molecule and a magnetic moment arises. This is $0.54 ~\mu_{\mathrm{B}}$ 
for the meta configuration [figure \ref{homo_deca}(a)] and $0.32 ~\mu_{\mathrm{B}}$ for the structure of figure \ref{homo_deca}(c). No moment is
then found for the two symmetric configurations [figure~\ref{homo_deca}(b) and figure~\ref{homo_deca}(d)]. 

\begin{figure}[ht]
  \begin{centering}
  \includegraphics[width=0.75\linewidth,clip=true]{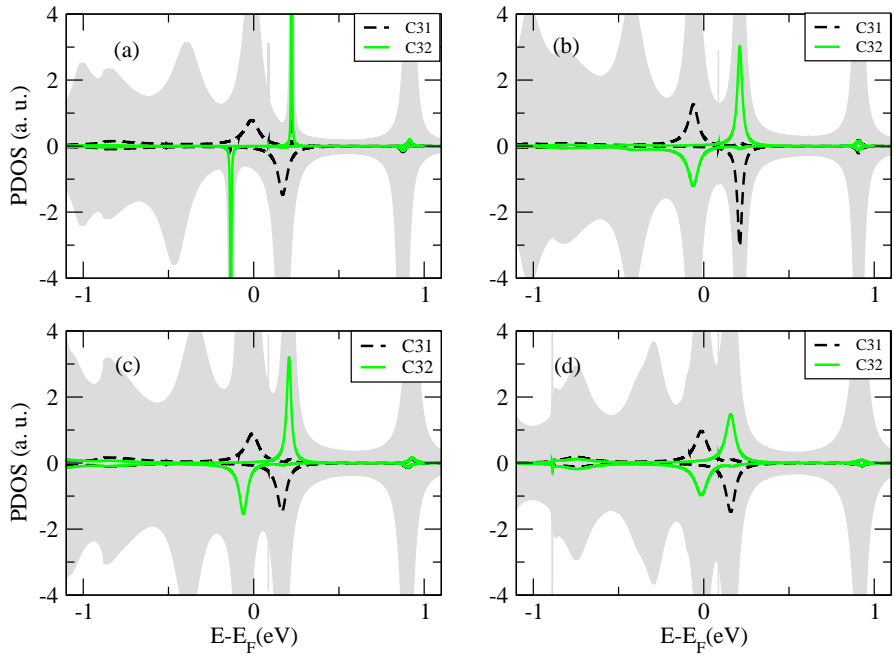}
  \caption{\label{pdos_deca}\small{PDOS for two carbons atom in the middle of decacene molecule (C31 and C32) for different anchoring 
  configurations: a) two links, meta; b) two links, para; c) three links d) four links. The grey shadow is the total DOS of the molecule.}}
\end{centering}
\end{figure}

\begin{figure}[ht]
  \begin{centering}
  \includegraphics[width=0.75\linewidth,clip=true]{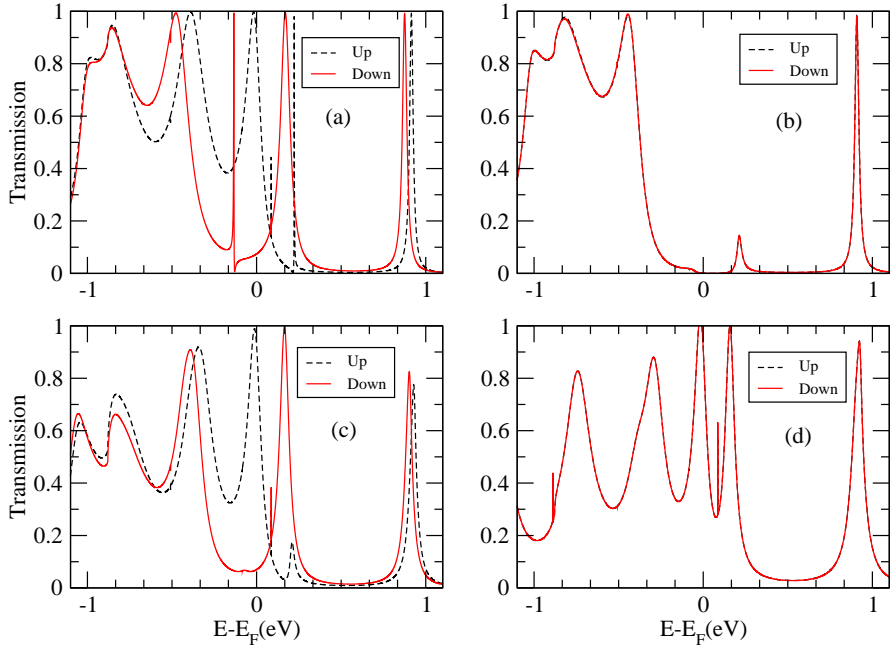}
  \caption{\label{trans_deca}\small{Transmission coefficient for decacene for the different anchoring configurations: a) two links, meta; 
  b) two links, para; c) three links d) four links. Up and Down indicate respectively the majority and minority spins contributions.}}
\end{centering}
\end{figure}

In order to analyze the effect of the different linking geometries on the electronic structure and the transport properties of the junctions,
we calculate for all the configurations the DOS projected on the central C atoms of decacene (figure \ref{pdos_deca}), and the zero-bias transmission 
coefficient (figure~\ref{trans_deca}). Note that in figure \ref{pdos_deca} we use no additional broadening, so that the width of the DOS
peaks is caused entirely by their coupling to the Au electrodes.
For the meta configuration [figure\ref{pdos_deca}(a)] the broadening of the DOS for states located on the upper edge of the molecule (atom C31) 
is larger then that of the states located on the lower edge (atom C32). This indicates strong coupling for the upper edge state and weak coupling, i.e. 
stronger localization, for the lower one, as expected from the bonding geometry. The spin-splitting of the states with higher coupling to the 
Au electrodes is reduced from its value for decacene in the gas phase. Furthermore a strong electron coupling enhances the partial charge 
transfer from the Au electrodes, so that the strongly coupled edge state is closer to the Fermi energy ($E_\mathrm{F}$) for both the HOMO and 
LUMO. Consequently the zero-bias transmission [figure ~\ref{trans_deca}(a)] for majority and minority spins is different. Intriguingly, at 
$E_\mathrm{F}$ the transmission coefficient for majority electrons is much larger than that for the minority ones, so that we expect a large 
spin-polarization of the current at small biases. This demonstrates that it is possible to obtain a spin-polarized current in $n$-acenes 
without using magnetic leads by simply exploiting their intrinsic magnetic nature and by engineering an asymmetric bonding to a metallic surface.

In the para configuration [figure~\ref{homo_deca}(b)] majority and minority states have the same coupling strength to the electrodes, since the 
two edges of the molecule are identically coupled. The PDOS [figure \ref{pdos_deca}(b)] associated to the two edges is symmetric and consequently 
no spin polarization is seen in the transmission coefficient of figure~\ref{trans_deca}(b). Then, when the molecule is anchored to Au through three links 
[figure ~\ref{homo_deca}(c)], although the minority spin density is not as localized as in the meta configuration, it is still more localized than the majority 
spin density. As such also here the electron transmission is different transmission for the two spins directions. Finally, when the molecule is connected
to both the edges with a total of 4 thiol groups we recover the complete symmetry between the two edges, and therefore the non-spin polarized 
transmission [see figures  \ref{pdos_deca}(d) and  \ref{trans_deca}(b)].
The coupling strength to the electrodes can be quantitatively described as indicated in \cite{magmol}. In brief, one can 
approximate the transmission coefficient as a function of energy, $T_\alpha(E)$, around a particular molecular orbital with energy 
$\epsilon_\alpha$ with a Breit-Wigner-like expression 
\begin{equation*}
T_\alpha(E)=\frac{\gamma_\alpha^\mathrm{L}\gamma_\alpha^\mathrm{R}}{(E-\epsilon_\alpha)^2+(\gamma_\alpha/2)^2}\:.
\end{equation*}
Here $\gamma_\alpha^\mathrm{L}$ ($\gamma_\alpha^\mathrm{R}$) is the coupling strength of that particular molecular orbital with the 
left-hand side (right-hand side) electrode. As such the full width at half maximum (FWHM) of a peak in the transmission coefficient is 
$\mathrm{FWHM}=\gamma_\alpha=\gamma_\alpha^\mathrm{L}+\gamma_\alpha^\mathrm{R}$. For symmetric coupling to the electrodes 
we have $\gamma_\alpha^\mathrm{L}=\gamma_\alpha^\mathrm{R}=\mathrm{FWHM}/2$. Note that the same level broadening is
found for the DOS and that this can be directly extracted from the NEGF density matrix.

Let us focus as an example on the meta configuration. In this case the majority spin levels, localized on the upper edge of the molecule,
are coupled to the electrodes via the thiol group. In contrast the minority spin levels, localized on the lower edge, can be accessed from the
electrodes only through tunneling in vacuum, since there is no direct binding atom. We then calculate a FWHM for the majority HOMO and LUMO
respectively of 100~meV and 67~meV. Although minor differences in the orbital wave-funciton produce differences in the FWHM we conclude
that the overall coupling of the upper edge levels is of the order of $\gamma_\alpha=84$~meV, which corresponds to a life-time of 5~ps.
The same analysis carried out for the minority spins gives us $\gamma_\alpha=2$~meV and a life-time of 200~ps. We then conclude that
the electronic coupling of the thiolated edge is approximately 40 times stronger than that of the non-thiolated one. 

We now briefly discuss the potential consequences arising from the LDA underestimation of the HOMO-LUMO gap. In general the main 
drawback is that both the HOMO and the LUMO are artificially too close to the electrodes $E_\mathrm{F}$. Therefore the calculated charge 
transfer from the molecule majority HOMO to Au is likely to be overestimated. As a consequence also the net magnetic moment is probably
overestimated. If one assumes that the charge transfer is negligible at zero-bias then the total magnetic moment will be close to zero. 
However even in such a situation a charge transfer and consequently a net magnetic moment can originate either at finite bias or by applying 
an electrostatic gate to the molecules. Therefore the finding of a total magnetic moment on the molecule is a general feature of such an 
interface, since it depends entirely on the fact that there is a considerably different electronic coupling to Au for the two spin channels of the
molecule. The fact that such a magnetic moment that can be potentially tuned via bias or gate voltage opens intriguing possibilities for device 
applications.

\begin{figure}[ht]
  \begin{centering}
  \includegraphics[width=0.55\linewidth,clip=true]{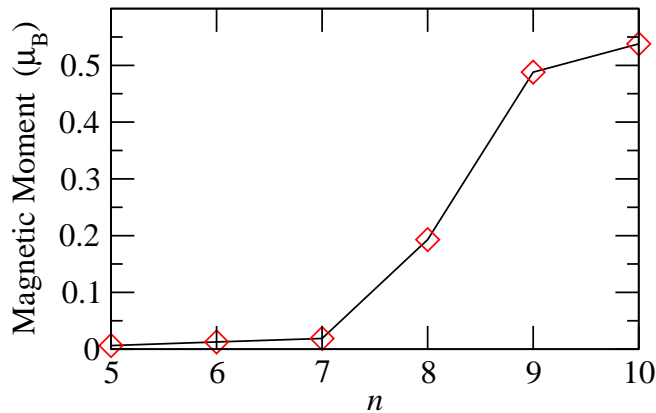}
  \caption{\label{spin_polarization}\small{Magnetic moment of the Au/$n$-acene/Au junction as a function of the molecule length, $n$.}}
\end{centering}
\end{figure}

\begin{figure}[ht]
  \begin{centering}
  \includegraphics[width=0.75\linewidth,clip=true]{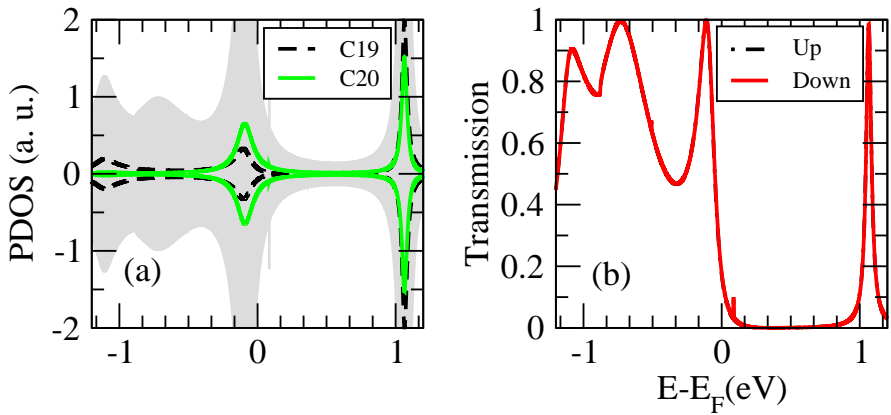}
  \caption{\label{PDOS_penta}\small{Projected density of states (a) and transmission coefficient (b) for pentacene in the 
  meta anchoring configuration.}}
\end{centering}
\end{figure}

In order to compare results for molecules of different length, we performed additional calculations for the Au/$n$-acene/Au junction in the
meta configuration and for different $n$. In general we find spin-polarized transmission for molecules made up of more than 8 rings 
(see figure~\ref{spin_polarization} for the magnetic moment of the junction as a function of $n$). The features of these magnetic
molecule junctions are qualitatively similar to those described in details for decacene. For the sake of comparison in figure~\ref{PDOS_penta} 
we present the PDOS and the zero-bias transmission for a representative non-magnetic acene, namely pentacene. Since the ground state is non-magnetic
the PDOS of the carbon atoms on the upper edge is the same as that on the lower edge. Therefore we find a completely non spin-polarized 
zero-bias transmission [see figure~\ref{PDOS_penta}(b)]. Importantly transport measurements for pentacene on Au indicate hole conductance~\cite{Jurchesco}. 
This is consistent with our results, which do indicate HOMO transport. However due to the underestimated IP our low bias conductance is likely to 
be overestimated.

\section{Conclusions}

We have examined the electronic structure and spin transport properties of higher acenes molecules attached to Au electrodes. These have
been compared with those of the smaller members of the acene family. We find that the isolated molecules have a spin-polarized ground state 
for molecules containing more then 8 acene rings. However the total moment vanishes even for the magnetic acenes, since the electrons in the 
two degenerate HOMO levels couple anti-ferromagnetically to each other. These results are in good agreement with both previous calculations 
and with the most recent experimental evidence. When such molecules are attached to Au via thiol linkers, a transition from a non-magnetic to 
a magnetic ground state occurs at 8 acene rings. The coupling between the two HOMO levels is still found to be anti-ferromagnetic. However, 
when the electronic interaction of the molecule with the electrodes is different between the two molecule edges than fractional charge transfer
leads to a total net spin. The Au/$n$-acene/Au system acts then as a spin filter. Such a spin-filtering property can be achieved by binding the 
molecule to Au via thiol linkers in an asymmetric way. We finally speculate on the fact that the charging state of the molecule and hence its magnetic
moment can be controlled by bias and/or electrostatic gating. 

\ack
IR, TA and SS acknowledge financial support from Science foundation of Ireland (07/IN.1/I945 and No. 07/RFP/PHYF235), CRANN and the
EU FP7 project ATHENA. Computational resources have been provided by the HEA IITAC project managed by TCHPC and by ICHEC.

\end{document}